\documentclass[a4paper,11pt]{article}
\usepackage{pos}

\newcommand{\bea}{\begin{eqnarray}} 
\newcommand{\eea}{\end{eqnarray}}

\newcommand{\JN}{J/\psi N}

\title{The $J/\psi$-nucleon interaction mechanism: A theoretical study based on scattering length }

\author*[a,b,c]{Bing Wu}
\author[d]{Xiang-Kun Dong}
\author[a]{Meng-Lin Du}
\author[b,c,e,f]{Feng-Kun Guo}
\author[g,b,c,e,f]{Bing-Song~Zou}

\affiliation[a]{School of Physics, University of Electronic Science and Technology of China,\\
 Chengdu 611731, China}

\affiliation[b]{CAS Key Laboratory of Theoretical Physics, Institute of Theoretical Physics,\\ Chinese Academy of Sciences, Beijing 100190, China}

\affiliation[c]{School of Physical Sciences, University of Chinese Academy of Sciences,\\ Beijing 100049, China}

\affiliation[d]{Helmholtz-Institut f\"{u}r Strahlen- und Kernphysik and Bethe Center for
Theoretical Physics, \\Universit\"{a}t~Bonn,  D-53115~Bonn,~Germany}

\affiliation[e]{Peng Huanwu Collaborative Center for Research and Education, Beihang University,\\ Beijing 100191, China}

\affiliation[f]{Southern Center for Nuclear-Science Theory (SCNT), Institute of Modern Physics,\\ Chinese Academy of Sciences, Huizhou 516000, China}

\affiliation[g]{Department of Physics, Tsinghua University,\\ Beijing 100084, China}

\emailAdd{wu.bing@uestc.edu.cn}
\emailAdd{xiangkun@hiskp.uni-bonn.de}
\emailAdd{du.ml@uestc.edu.cn}
\emailAdd{fkguo@itp.ac.cn}
\emailAdd{zoubs@mail.tsinghua.edu.cn}

\abstract{The low-energy $J/\psi N$ scattering is of significant importance for various reasons. It is deeply interconnected with the hidden-charm $P_c$ pentaquark states, provides insights into the role of gluons in nucleon structures, and is pertinent to the properties of $J/\psi$ in nuclear medium. The scattering can occur through two distinct mechanisms: the coupled-channel mechanism involving open-charm meson-baryon intermediate states $\Lambda_c \bar D^{(*)}$ and $ \Sigma_c^{(*)}\bar D^{(*)}$, and the soft-gluon exchange mechanism. In this study, we investigate the $S$-wave $J/\psi N$ scattering length arising from both mechanisms. Our findings indicate that both mechanisms lead to attractive interactions, yielding scattering lengths of $[-10, -0.1] \times 10^{-3}$ fm for the coupled-channel mechanism and $<-0.16$ fm for the soft-gluon exchange mechanism, respectively. Notably, the soft-gluon exchange mechanism produces a scattering length that is at least one order of magnitude larger than that from the coupled-channel mechanism, indicating its predominance. These findings can be corroborated through lattice calculations and will enhance our understanding of scattering processes that violate the Okubo-Zweig-Iizuka rule.}

\FullConference{The 11th International Workshop on Chiral Dynamics (CD2024)\\
 26-30 August 2024\\
Ruhr University Bochum, Germany\\}

\begin{document}
\maketitle

\section{Introduction}
The low-energy $J/\psi N$ interaction has attracted sustained interest for a long time~\cite{Peskin:1979va}, as it provides vital insights into the role of gluons in hadron structures and interactions among hadrons. These remain challenging problems in quantum chromodynamics (QCD) within the nonperturbative region. This interaction plays a crucial role in describing the photo- and hadro-production of charmonia on nuclear targets, a topic that has garnered increasing interest, especially after the discovery of pentaquark $P_c$ states by the LHCb Collaboration~\cite{LHCb:2015yax,LHCb:2019kea}. Additionally, it provides a possible way to probe the trace anomaly contribution to the nucleon mass~\cite{Kharzeev:1995ij,Kharzeev:1998bz}. It is also relevant to the properties of the $J/\psi$ in nuclear medium~\cite{Sibirtsev:2005ex} and to identifying signals of the quark-gluon plasma~\cite{Barnes:2003vt}. These processes will be investigated in future experiments at the Electron-Ion Collider (EIC)~\cite{Burkert:2022hjz}, the potential 22~GeV upgrade at Jefferson Laboratory~\cite{Accardi:2023chb}, and the Electron-Ion Collider in China~\cite{Anderle:2021wcy}.

Since the $J/\psi$ meson and nucleon do not share any valence quarks, their interaction via meson exchange is suppressed, in accordance with the Okubo–Zweig–Iizuka (OZI) rule~\cite{Okubo:1963fa,Iizuka:1966fk}. The $J/\psi N$ scattering, and more generally OZI-suppressed scattering processes, can proceed through two distinct mechanisms. The first mechanism involves the exchange of multiple gluons, which consists of two key components: (1) the charmonium-gluon coupling, described by the chromopolarizability, which characterizes the ability of the charmonium to emit gluons~\cite{Chen:2019hmz}, and (2) the matrix elements of the gluonic operator between single nucleon states, associated with the trace anomaly contribution to the nucleon mass~\cite{Chanowitz:1972vd,Crewther:1972kn,Voloshin:1980zf}. The second mechanism encompasses coupled channels arising from rescattering from the $J/\psi N$ state to channels involving open-charm meson and baryon states, specifically $\Lambda_c \bar D^{(*)}$ and $\Sigma_c^{(*)}\bar D^{(*)}$~\cite{Du:2020bqj}, and back to $J/\psi N$. Therefore, a study of the trace anomaly contribution to the nucleon mass using $J/\psi$ as a probe is feasible only if the first mechanism dominates. Deciphering the mechanism of low-energy $J/\psi N$ scattering is crucial for this purpose.

Despite extensive exploration of the $J/\psi N$ interaction using various methods, the strength of this interaction remains a topic of ongoing debate. Different theoretical studies report significant variations in the $S$-wave $J/\psi N$ scattering length, spanning several orders of magnitude~\cite{Brodsky:1997gh,Sibirtsev:2005ex,Du:2020bqj,Pentchev:2020kao}. In particular, lattice QCD studies have not yielded consistent results, with predictions ranging from a strong attraction~\cite{Beane:2014sda} to a very weak $J/\psi N$ interaction~\cite{Skerbis:2018lew}.

One ongoing effort is to extract the $J/\psi N$ scattering length from the near-threshold photoproduction of $J/\psi$ on the proton~\cite{Strakovsky:2019bev,Pentchev:2020kao,GlueX:2023pev}. This method is based on the vector-meson dominance model, where a nearly on-shell photon is converted into a $J/\psi$, which then interacts with the proton~\cite{Kharzeev:1998bz}. However, the $J/\psi N$ scattering length is defined for on-shell $J/\psi$ and nucleon, while the intermediate $J/\psi$ connected to the photon is highly off-shell. Additionally, the applicability of the vector-meson dominance model in the charmonium sector for this process has been challenged in Refs.~\cite{Du:2020bqj,Xu:2021mju,JointPhysicsAnalysisCenter:2023qgg}. In particular, it has been demonstrated in Refs.~\cite{Du:2020bqj,JointPhysicsAnalysisCenter:2023qgg} that the $J/\psi$ photoproduction data can be well described by the coupled-channel mechanism, and the predicted existence of $\Lambda_c\bar D^{(*)}$ threshold cusps~\cite{Du:2020bqj} is consistent with the GlueX data~\cite{GlueX:2023pev}.

\section{Coupled-channel mechanism}
In the coupled-channel mechanism, the interaction between $J/\psi$ and the nucleon is facilitated by open-charm meson-baryon intermediate states. The scattering amplitudes for the coupled-channel system, which includes $J/\psi N$, $\eta_c N$, $\Lambda_c\bar D^{(*)}$, and $\Sigma_c^{(*)}\bar D^{(*)}$, have been constructed by solving the Lippmann-Schwinger equation within the framework of chiral effective field theory, as detailed in Ref.~\cite{Du:2021fmf}. The long-range interaction is mediated by one-pion exchange, while the short-range coupled-channel interactions are parameterized through contact terms. Next-to-leading-order contact terms for the transitions from $S$-wave to $D$-wave are included to ensure cutoff independence of the results. Heavy quark spin symmetry is utilized to relate various couplings involving charmed hadrons that belong to the same spin multiplet. The parameters are determined by fitting to the LHCb data on the $J/\psi p$ invariant mass distribution from the decay $\Lambda_b^0 \rightarrow J/\psi p K^{-}$~\cite{LHCb:2019kea}. Further details of the framework can be found in Ref.~\cite{Du:2021fmf}. Using the scattering amplitudes obtained therein, we calculate the $S$-wave $J/\psi N$ scattering length, defined through the effective range expansion as
\begin{equation}
    p \cot\delta_{0,J/\psi N} = -\frac{1}{a_{0,J/\psi N}} + \frac{1}{2} r_{0,J/\psi N}\, p^2 + \cdots,
\end{equation}
where $p$ represents the magnitude of the $J/\psi$ momentum in the $J/\psi N$ c.m. frame. Here, $\delta_{0,J/\psi N}$, $a_{0,J/\psi N}$, and $r_{0,J/\psi N}$ denote the $S$-wave phase shift, scattering length, and effective range, respectively. The result is given by
\begin{align}
    a_{0,J/\psi N}^\text{CC} \in [-10, -0.1] \times 10^{-3}~\text{fm},
    \label{eq:acc}
\end{align}
where the significant uncertainty arises from the fitting of parameters to the LHCb data. We use "CC" to denote the coupled-channel mechanism. This result indicates that the charmed meson-baryon coupled channels lead to a very weak attraction between the $J/\psi$ and the nucleon.

\section{Gluon exchange mechanism}
\begin{figure}[tb]
\centering
    \includegraphics[width=0.6\linewidth]{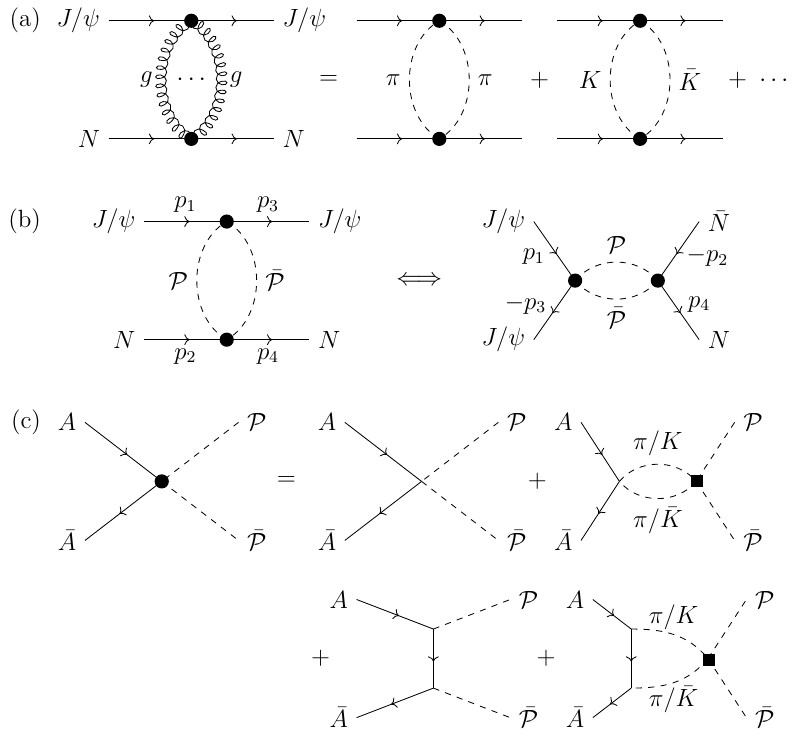}
\caption{(a) Soft-gluon exchange between $J/\psi$ and the nucleon, which is effectively equivalent to the exchange of $\pi\pi$, $K\bar{K}$, and other heavier hadrons. (b) Crossing symmetry relation between $\JN$ scattering and the process $J/\psi J/\psi \to N\bar{N}$. (c) Amplitude for $A\bar{A} \to \mathcal{P}\bar{\mathcal{P}}$ (where $A = N, J/\psi$ and $\mathcal{P} = \pi, K$) incorporating $\pi\pi$-$K\bar{K}$ coupled-channel rescattering. The black dot represents the full $A\bar{A} \to \mathcal{P}\bar{\mathcal{P}}$ amplitude, while the black square denotes the $\pi\pi$-$K\bar{K}$ coupled-channel rescattering contribution.}
  \label{fig:feyndiag}
\end{figure}

Because the exchanged gluons between $J/\psi$ and the nucleon must be in a color-singlet and isospin-singlet combination, their effects can be captured by inserting a complete set of hadrons with the same quantum numbers as the exchanged multiple gluons. The most important contribution comes from the lightest $\pi\pi$ exchange, as well as the $K\bar{K}$ exchange due to the strong $\pi\pi$-$K\bar{K}$ coupling, as depicted in Fig.~\ref{fig:feyndiag}~(a).

The exchange of soft gluons with the vacuum quantum numbers can be evaluated using the technique of dispersion relations. This approach has been employed to demonstrate that the exchange of correlated $S$-wave $\pi\pi$ can be parameterized in terms of a scalar sigma exchange~\cite{Donoghue:2006rg,Wu:2023uva}. The $J/\psi N$ scattering amplitude from the $\pi\pi$ and $K\bar{K}$ (denoted by $\mathcal{P}\bar{\mathcal{P}}$ in the following, with $\mathcal{P} = \pi, K$) exchanges is related to the amplitude for $J/\psi J/\psi \to \mathcal{P}\bar{\mathcal{P}} \to N\bar{N}$ by crossing symmetry, as shown in Fig.~\ref{fig:feyndiag} (b). The latter can be calculated through a dispersion relation as~\cite{Wu:2024xwy}
\begin{align}
    \mathcal{T}_{0,J/\psi J/\psi \to N(\lambda_3)\bar{N}(\lambda_4)}(s) = \sum_{\mathcal{P}=\pi,K} \frac{\lambda_3 + \lambda_4}{\pi} \int_{4M_\pi^2}^{+\infty} {\rm d}s' \frac{T_{0,J/\psi J/\psi \to \mathcal{P}\bar{\mathcal{P}}}(s') \, \rho_{\mathcal{P}}(s') \, \mathcal{T}_{0,N\bar{N} \to \mathcal{P}\bar{\mathcal{P}}}^{*}(s')}{s' - s - i\epsilon}, \label{TJpsiJpsitoNbarNinmaintext}
\end{align}
where $s$ represents the square of the total energy of $\mathcal{P}\bar{\mathcal{P}}$ in their center-of-mass (c.m.) frame, $\lambda_3$ ($\lambda_4$) denotes the helicity of $N$ ($\bar{N}$), and 
\begin{align}
\rho_{\mathcal{P}}(s) = \theta\left(\sqrt{s} - 2M_\mathcal{P}\right) \frac{\sqrt{1 - 4M_\mathcal{P}^2/s}}{16\pi}
\end{align}
is the phase space factor of $\mathcal{P}\bar{\mathcal{P}}$, with $\theta(s)$ being the Heaviside step function. Here, $T_{0,A\bar{A} \to \mathcal{P}\bar{\mathcal{P}}}(s)$ is the scattering amplitude for the process $A\bar{A} \to \mathcal{P}\bar{\mathcal{P}}$, where $A = N$ or $J/\psi$. The quantity 
\begin{align}    
\mathcal{T}_{0,N\bar{N} \to \mathcal{P}\bar{\mathcal{P}}}(s) = \frac{T_{0,N\bar{N} \to \mathcal{P}\bar{\mathcal{P}}}(s)}{\sqrt{s - 4m_N^2}}
\end{align}
is introduced to avoid the kinematic singularity arising from the partial-wave projection~\cite{Martin:1970hmp,Wu:2023uva}. The subscript $0$ indicates that $\mathcal{P}\bar{\mathcal{P}}$ is in an $S$-wave; since this is always the case, this subscript will be omitted in the following for simplicity.

\begin{figure}[tb]
\centering
    \includegraphics[width=0.6\linewidth]{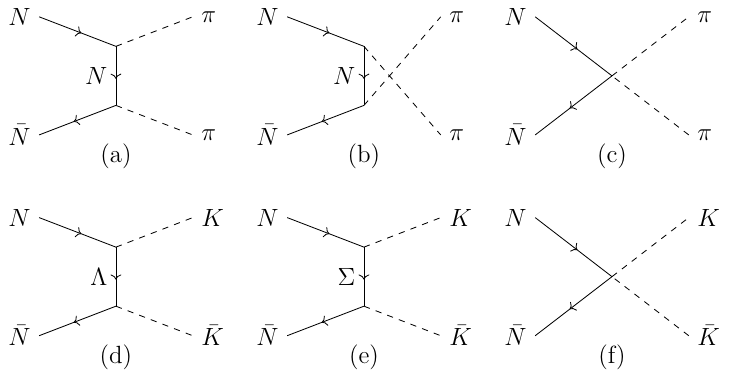}
\caption{Tree-level Feynman diagrams for the $N\bar{N}\to\pi\pi$ and $N\bar{N}\to K\bar{K}$ processes.}
  \label{FIG2}
\end{figure}

We include the interaction in the $\pi\pi$-$K\bar{K}$ coupled-channel system, as shown in Fig.~\ref{fig:feyndiag}~(c), which is significant and contains the influences of scalar $\sigma$ and $f_0(980)$ exchanges. This is achieved using the Muskhelishvili-Omn\`es representation~\cite{Hoferichter:2012wf,Lin:2022dyu}. The procedure has been outlined in Ref.~\cite{Dong:2021lkh}. We have
\begin{align}
\vec{\mathcal{M}}_{A\bar{A}}(s) = \vec{L}_{A\bar{A}}(s) + \Omega(s) \Biggl[ \vec{P}^{(n-1)}_{A\bar{A}}(s) - \frac{s^n}{\pi} \int_{4M_\pi^2}^{+\infty} {\rm{d}}z \frac{{\rm Im} \left[ \Omega^{-1}(z) \right]}{(z - s)z^n} \Theta(z) \vec{L}_{A\bar{A}}(z) \Biggr], \label{equMOrepresentation}
\end{align}
with
\begin{align*}
\vec{\mathcal{M}}_{A\bar{A}}(s) &= \left( \mathcal{M}_{A\bar{A} \to \pi\pi}(s), \mathcal{M}_{A\bar{A} \to K\bar{K}}(s) \right)^{\rm T}, \\
\vec{L}_{A\bar{A}}(s) &= \left( L_{A\bar{A} \to \pi\pi}(s), L_{A\bar{A} \to K\bar{K}}(s) \right)^{\rm T}, \\
\vec{P}^{(n-1)}_{A\bar{A}}(s) &= \left( P^{(n-1)}_{A\bar{A} \to \pi\pi}(s), P^{(n-1)}_{A\bar{A} \to K\bar{K}}(s) \right)^{\rm T}, \\
\Theta(s) &= {\rm diag} \left( \theta\left(s - 4M_\pi^2\right), \theta\left(s - 4M_K^2\right) \right),
\end{align*}
where the elements of $\vec{P}^{(n-1)}(s)$ are polynomials of order $n-1$, and $\vec{L}_{A\bar{A}}(s)$ represents the possible contributions that do not have the right-hand cuts due to the on-shellness of the intermediate meson pairs (they can have left-hand cuts). $\Omega(s)$ is the Omn\`es matrix for the $S$-wave isoscalar $\mathcal{P}\bar{\mathcal{P}}$ coupled-channel interaction. We employ the Omn\`es matrix elements extracted in Ref.~\cite{Ropertz:2018stk}, which have been calibrated to align with the $\pi\pi$-$K\bar{K}$ scattering amplitudes reported in Ref.~\cite{Dai:2014zta} at lower energies. Here, $\mathcal{M}_{A\bar{A} \to \mathcal{P}\bar{\mathcal{P}}}(s) = T_{J/\psi J/\psi \to \mathcal{P}\bar{\mathcal{P}}}(s)$ for $J/\psi J/\psi \to \mathcal{P}\bar{\mathcal{P}}$ and $\mathcal{M}_{A\bar{A} \to \mathcal{P}\bar{\mathcal{P}}}(s) = \mathcal{T}_{N \bar{N} \to \mathcal{P}\bar{\mathcal{P}}}(s)$ for $N\bar{N} \to \mathcal{P}\bar{\mathcal{P}}$.

The amplitudes $\vec{L}_{A\bar{A}}(s)$, which receive left-hand cut contributions, and the polynomials $\vec{P}^{(n-1)}_{A\bar{A}}(s)$ can be determined by matching to the tree-level amplitudes derived from chiral Lagrangians. We employ chiral Lagrangians up to $\mathcal{O}(p^2)$, where $p$ represents a small momentum scale. Specifically, we utilize the baryon-meson chiral Lagrangian up to the next-to-leading order (NLO)~\cite{Oller:2006yh}, as well as the leading order (LO) chiral Lagrangian for the emission of light pseudoscalar mesons from charmonia~\cite{Chen:2015jgl}.

For $A = N$, the left-hand-cut part $L_{N\bar{N} \to \mathcal{P}\bar{\mathcal{P}}}(s)$ corresponds to the processes illustrated in Fig.~\ref{FIG2}~(a), (b), (d), and (e), where the involved meson-baryon couplings are derived from the LO chiral Lagrangian. The polynomials are matched to the tree-level amplitudes $\mathcal{A}_{N\bar{N} \to \mathcal{P}\bar{\mathcal{P}}}(s)$ of the processes illustrated in Fig.~\ref{FIG2}~(c) and (f). The involved low-energy constants (LECs) are taken from Fit~II in Ref.~\cite{Ren:2012aj}. Furthermore, given that the explicit forms of $\mathcal{A}_{N\bar{N} \to \mathcal{P}\bar{\mathcal{P}}}(s)$ are linear polynomials in $s$, it is sufficient to adopt the twice-subtracted ($n=2$) dispersion relation~\cite{Wu:2023uva}. 

For $A = J/\psi$, there is no left-hand-cut contribution since no resonance near the $J/\psi\pi$ or $J/\psi K$ threshold exists. The polynomials are matched to the tree-level chiral amplitudes given by
\begin{align}
    \mathcal{A}_{J/\psi J/\psi \to \mathcal{P}\bar{\mathcal{P}}}(s) = -\frac{2I_{\mathcal{P}}}{F_\pi^2} \Biggl\{ & c_1^{(11)} \left(s - 2M_{\mathcal{P}}^2\right) + 2c_m^{(11)} M_{\mathcal{P}}^2 \notag \\
    &\ + \frac{c_2^{(11)}}{12M_{J/\psi}^2} \left[ s^2 + 2s(M_{\mathcal{P}}^2 + M_{J/\psi}^2) - 8M_{\mathcal{P}}^2 M_{J/\psi}^2 \right] \Biggr\}, \label{eq:AJpsiJpsi}
\end{align}
with $I_{\pi} = \sqrt{3/2}$ and $I_{K} = \sqrt{2}$ being the isospin factors. The LECs $c_{1,2,m}^{(11)}$ are related to the chromopolarizability of $J/\psi$, $\alpha^{(11)}$, as~\cite{Brambilla:2015rqa} 
\begin{align}
    c^{(ij)}_1 = -\left(4 + 3\kappa\right) f^{(ij)}, \ 
    c^{(ij)}_2 = 12 \kappa f^{(ij)}, \
    c^{(ij)}_m = -6f^{(ij)}, 
    \label{eq:cij}
\end{align}
where $f^{(ij)}={\pi^2\sqrt{m_{\psi_i} m_{\psi_j}}F_\pi^2}\alpha^{(ij)}/{\beta_0}$, $\beta_0=({11N_c}-{2 N_f})/{3}$ is the first coefficient of the QCD beta function with $N_c$ and $N_f$ being the numbers of colors and light-quark flavors, $F_\pi$ is the pion decay constant, and $\kappa$ is an unknown parameter. The upper index $(ij)$ is introduced to represent $\psi(iS)$ and $\psi(jS)$ states for the involved charmonia (hence $(11)$ appears in Eq.~\eqref{eq:AJpsiJpsi}).

\begin{figure}[tbh]
    \centering
    \includegraphics[width=0.6\linewidth]{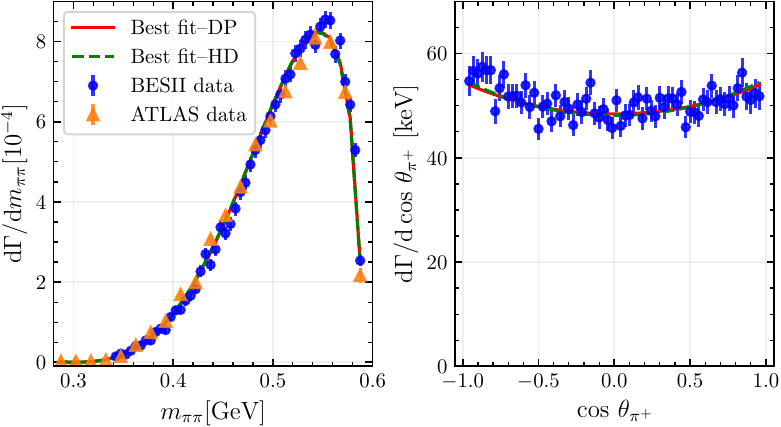}
    \caption{Fit to the BESII data~\cite{BES:2006eer} and ATLAS data~\cite{ATLAS:2016kwu} for the $\psi(2S) \to J/\psi \pi^+\pi^-$ transition: dipion invariant mass distribution (left) and helicity angular distribution (right). The labels ``Best fit--DP'' and ``Best fit--HD'' correspond to the fitting results obtained using the Omn\`es matrix from Ref.~\cite{Ropertz:2018stk} and Ref.~\cite{Hoferichter:2012wf}, respectively.}
    \label{fig:fitall}
\end{figure}
Although the value of the chromopolarizability for the $J/\psi$ determined using various methods~\cite{Brambilla:2015rqa,Dong:2022rwr} bears a quite large uncertainty, the off-diagonal chromopolarizability $\alpha^{(21)}$ can be well fixed using experimental data for the $\psi(2S) \to J/\psi \pi^+ \pi^-$ decay~\cite{BES:2006eer,ATLAS:2016kwu}. Determinations considering the $\pi\pi$ final-state interactions for this transition have been performed in Refs.~\cite{Guo:2006ya,Chen:2019hmz,Dong:2021lkh}. Using the Omn\`es matrix from Refs.~\cite{Ropertz:2018stk,Hoferichter:2012wf}, an updated fit to the data of the $\pi\pi$ invariant mass distribution and the helicity angular distribution for the $\psi(2S) \to J/\psi \pi^+ \pi^-$ decay~\cite{BES:2006eer,ATLAS:2016kwu} leads to $\kappa = 0.26 \pm 0.01 \pm 0.01$ and
\begin{align}
    |\alpha^{(21)}| = (1.18 \pm 0.01 \pm 0.05) \ \text{GeV}^{-3},
\end{align}
where the first uncertainties come from the errors of the experimental data, while the second ones represent the differences arising from the Omn\`es matrices in Refs.~\cite{Ropertz:2018stk,Hoferichter:2012wf}. A comparison of the fit results with the data is shown in Fig.~\ref{fig:fitall}.

Building upon Eq.~(\ref{equMOrepresentation}) and employing the dispersion relation~\eqref{TJpsiJpsitoNbarNinmaintext}, we derive the $s$-channel helicity amplitude $\mathcal{T}_{J/\psi J/\psi \to N(\lambda_3)\bar{N}(\lambda_4)}(s)$ for the process $J/\psi J/\psi \to N\bar{N}$. A Gaussian form factor, $\exp\left[-(s' - s)/\Lambda^2\right]$, is introduced into the integrand of the dispersion integral in Eq.~\eqref{TJpsiJpsitoNbarNinmaintext} to regularize the ultraviolet linear divergence. This specific form ensures that the long-distance behavior of the potential remains unaffected by the form factor~\cite{Reinert:2017usi}. Using the crossing relation, this framework enables us to construct the corresponding $t$-channel $S$-wave helicity amplitude for the process $J/\psi N \to J/\psi N$, as depicted in Fig.~\ref{fig:feyndiag} (b). From this, one can obtain the $S$-wave $J/\psi N$ scattering amplitude.

\section{$J/\psi$-nucleon scattering lengths}
We denote the $J/\psi N$ invariant mass squared as $t$. According to the exact crossing relation satisfied by the helicity amplitude of the system with spin and the relation between the helicity amplitude and the $L$-$S$ amplitude~\cite{Martin:1970hmp,Wu:2023uva}, the $S$-wave $J/\psi N$ scattering amplitude for definite total spin ($S$) can then be expressed as
\begin{align}
     T_{0,J/\psi N}(t)=&\sum\limits_{\mu_1\mu_2}\sum\limits_{\mu_1'\mu_2'}\frac{1}{2S+1}C_{1\mu_1';\frac{1}{2}(-\mu_2')}^{S(\mu_1'-\mu_2')}C_{1\mu_1;\frac{1}{2}(-\mu_2)}^{S(\mu_1-\mu_2)}\int\frac{{\rm d}\Omega}{4\pi} d^S_{\mu,\mu'}(\theta)\epsilon(\mu_1,0,0)\cdot\epsilon^*(\mu_1',\theta,0)\notag\\
    &\times e^{-i\pi \mu_2'}(-1)^{\mu_2-\mu_2'}\sum\limits_{\lambda_3'\lambda_4'}d^{\frac{1}{2}}_{\mu_2\lambda_3'}(\omega)d^{\frac{1}{2}}_{\mu_2'\lambda_4'}(\omega+\pi)(\lambda_3'+\lambda_4')T_{0,J/\psi J/\psi \to \bar{N}(\frac{1}{2})N(\frac{1}{2})}(s)\ ,\label{intpotinapp1}
\end{align}
where $T_{0,J/\psi J/\psi \to \bar{N}N}=\sqrt{s-4m_N^2}\mathcal{T}_{0,J/\psi J/\psi \to \bar{N}N}$, $\mu^{(\prime)}$ and $\lambda^{(\prime)}$ represent the helicities of corresponding partilces, $\epsilon(\lambda,\theta,\phi)$ denotes the polarization vector of $J/\psi$, $(\theta,\phi)$ represent the polar and azimuthal angles of the momentum of $J/\psi$ in the $J/\psi N$ c.m. frame. We choose the relative momentum of the initial two particles along the $z$-axis and the relative momentum of the final two particles within the $x$-$z$ plane. The Wigner rotation angle $\omega$ corresponds to the Lorentz transformation from the $s$-channel c.m. frame to the $t$-channel c.m. frame. The phase factor $e^{-i\pi \mu_2'}(-1)^{\mu_2-\mu_2'}$ is incorporated to ensure that the amplitude resulting from the crossing relation conforms to the correct phase convention, which can be determined through examining specific $s$-channel and $t$-channel amplitude instances. $C_{S_1S_{1z};S_2S_{2z}}^{SS_z}$ is the Clebsch-Gordan coefficient of SU(2).

In the $J/\psi N$ c.m. frame, we have $t = \left(\sqrt{|\vec{p}|^2 + M_{J/\psi}^2} + \sqrt{|\vec{p}|^2 + m_N^2} \right)^2$ and $s = -2|\vec{p}|^2(1 - \cos{\theta})$, where $|\vec{p}|$ denotes the magnitude of the relative momentum. At the threshold $t = (M_{J/\psi} + m_N)^2$, $s = 0$. Considering the limit 
\begin{align}
\lim_{|\vec{p}|\to0}\sum\limits_{\lambda_3'\lambda_4'}d^{\frac{1}{2}}_{\mu_2\lambda_3'}(\omega)d^{\frac{1}{2}}_{\mu_2'\lambda_4'}(\omega+\pi)(\lambda_3'+\lambda_4')=
\left\{
\begin{array}{cl}
 -\frac{\sin{\theta}}{\sqrt{2-2\cos{\theta}}} & {\rm if}\ \mu_2=\mu_2'=-\frac{1}{2}
\\
\sin\left(\theta/2\right) & {\rm if}\ \mu_2=\pm\frac{1}{2},\mu_2'=\mp\frac{1}{2}
\\
\frac{\sin{\theta}}{\sqrt{2-2\cos{\theta}}} & {\rm if}\ \mu_2=\mu_2'=\frac{1}{2}
\end{array}
\right. \label{ddomegainapp}
\end{align}
and the explicit form of polarization vectors in the helicity representation, it can be shown that at the $J/\psi N$ threshold $t_{\rm th} = (M_{J/\psi} + m_N)^2$, the $S$-wave $J/\psi N$ scattering amplitude satisfies
\begin{align}
    T_{0,J/\psi N}(t = t_{\rm th}) = 2m_N \mathcal{T}_{J/\psi J/\psi \to N(\frac{1}{2})\bar{N}(\frac{1}{2})}(s = 0), \label{potential2}
\end{align}
which is independent of the total spin $S$. The dimensional factor $2m_N$ arises from our normalization of the amplitudes in Eq.~\eqref{TJpsiJpsitoNbarNinmaintext}. The scattering length is given by
\begin{align}
a_{J/\psi N} = -\frac{T_{0,J/\psi N}(t = t_{\rm th})}{8\pi \sqrt{t_{\rm th}}} \ .
\end{align}

\begin{figure}[tb]
\centering
    \includegraphics[width=0.45\linewidth]{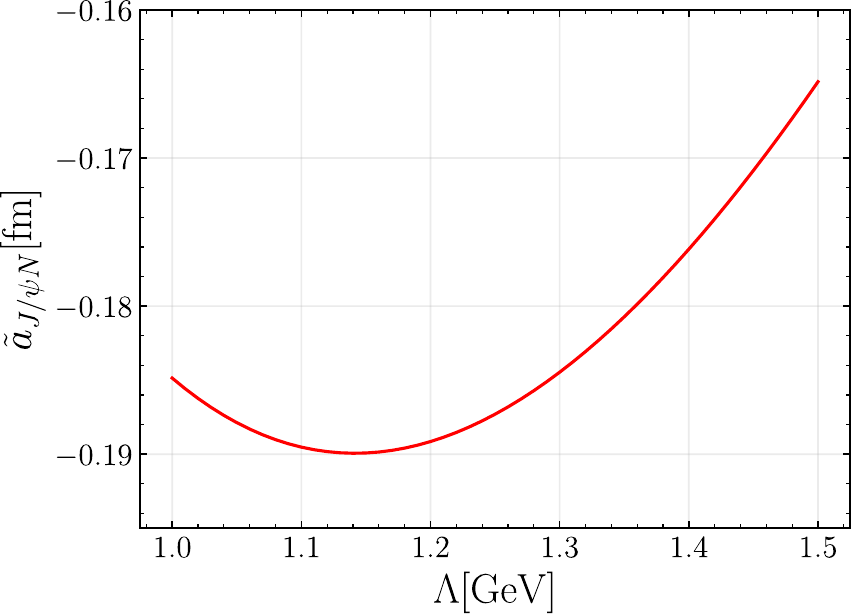}
\caption{Dependence of the $S$-wave $J/\psi N$ scattering length on the cutoff $\Lambda$, evaluated assuming $c^{(11)}_{1,2,m}=c^{(21)}_{1,2,m}$. 
}
  \label{scattringlength}
\end{figure}

To estimate the $J/\psi N$ scattering length, let us first take $c_{1,2,m}^{(11)} = c_{1,2,m}^{(21)}$ to evaluate $\mathcal{A}_{J/\psi J/\psi \to \mathcal{P}\bar{\mathcal{P}}}$ and denote the resulting scattering length as $\tilde{a}_{J/\psi N}$.\footnote{The sign of $\alpha^{(11)}$ is chosen to be positive, consistent with the estimate for considering the quarkonium as a pure color Coulomb system~\cite{Peskin:1979va}.} The result is shown in Fig.~\ref{scattringlength}. We find that $\tilde{a}_{J/\psi N}$ falls in the range of $-0.16$ to $-0.19$ fm as $\Lambda$ varies from $1.0$ GeV to $1.5$ GeV. The negative sign indicates an attractive interaction. Replacing the $J/\psi$ mass with the $\psi(2S)$ mass in Eq.~\eqref{eq:AJpsiJpsi}, we obtain $\tilde{a}_{\psi(2S) N}$ in the range of $-0.14$ to $-0.17$ fm.

It has been noticed in Ref.~\cite{Dong:2021lkh} that the absolute value of $\alpha^{(11)}$ should be larger than that of the off-diagonal $\alpha^{(21)}$. Thus, we have 
\begin{equation}
    a_{J/\psi N} \lesssim -0.16~{\rm fm}\ ,
\end{equation}
whose absolute value is at least one order of magnitude larger than that from the coupled-channel mechanism given in Eq.~\eqref{eq:acc}. 

Indeed, since the LECs $c^{(ij)}_{1,2,m}$ are proportional to $\alpha_{ij}$, using the Schwartz inequality~\cite{Sibirtsev:2005ex},
\begin{align}
    \alpha^{(11)}\alpha^{(22)} \geq |\alpha^{(21)}|^2, \label{eq:chromopolarizability}
\end{align}
the product of $\tilde{a}_{J/\psi N}$ and $\tilde{a}_{\psi(2S) N}$ derived above in fact sets a lower bound for the product of the $J/\psi N$ and $\psi(2S) N$ scattering lengths, 
\begin{align}
    a_{J/\psi N} a_{\psi(2S) N} \geq \tilde{a}_{J/\psi N} \tilde{a}_{\psi(2S) N} \approx {(-0.15\ \rm{ fm}})^2.
\end{align}

A recent lattice QCD calculation~\cite{Lyu:2024ttm} presents the result for $a_{J/\psi N}$ is in the range of $-0.42$ to $-0.28$~fm, which is consistent with the findings reported in this work.

In summary, we have calculated the contributions to the $S$-wave $J/\psi N$ scattering length from two mechanisms: (1) charmed meson-baryon coupled channels, based on the effective field theory formalism in Ref.~\cite{Du:2021fmf}, and (2) soft-gluon exchanges, modeled as light meson exchanges using dispersion relations. Both mechanisms yield attractive interactions, with scattering lengths in the ranges $[-10, -0.1] \times 10^{-3}$~fm (coupled-channel) and $<-0.16$~fm (soft-gluon exchange). Our results demonstrate that the low-energy $J/\psi N$ scattering is dominated by the soft-gluon exchange mechanism, making it possible to extract the trace anomaly contribution to the nucleon mass from $J/\psi N$ scattering. These findings can be validated using lattice QCD and provide a foundation for future studies of the gluonic structure of the nucleon through low-energy $J/\psi N$ scattering.

\end{document}